\documentclass[pdflatex,iicol,sn-mathphys-num]{sn-jnl}
\usepackage[T1]{fontenc}
\usepackage[utf8]{inputenc}
\usepackage{amsmath,amssymb,amsthm,mathtools}
\usepackage{xcolor}
\usepackage{microtype}
\usepackage{booktabs,array}
\usepackage{hyperref}
\usepackage{enumitem}
\hypersetup{hidelinks,pdfauthor={Alexey Sergeevich Osipov and Alexander Mettalinovich Tishin},pdftitle={Geometric selection of a single Weyl sector from symplectic orientation and a local causal record}}
\setlist[itemize]{topsep=3pt,itemsep=1pt,leftmargin=*}
\setlist[enumerate]{topsep=3pt,itemsep=1pt,leftmargin=*}
\newtheorem{theorem}{Theorem}[section]
\newtheorem{lemma}[theorem]{Lemma}
\newtheorem{proposition}[theorem]{Proposition}

\theoremstyle{definition}

\theoremstyle{remark}

\DeclareMathOperator{\Spin}{Spin}
\DeclareMathOperator{\Hom}{Hom}
\DeclareMathOperator{\LieSpan}{LieSpan}
\DeclareMathOperator{\End}{End}
\DeclareMathOperator{\Alt}{Alt}
\DeclareMathOperator{\Span}{span}
\newcommand{\D}{\mathrm{d}}
\newcommand{\DG}{\mathrm{DG}}

\newcommand{\cL}{\mathcal{L}}
\newcommand{\cA}{\mathcal{A}}
\newcommand{\cO}{\mathcal{O}}

\newcommand{\cB}{\mathcal{B}}
\newcommand{\eps}{\varepsilon}
\newcommand{\gfive}{\gamma^5_{\DG}}
\newcommand{\Jax}{\Xi_{\DG}}
\newcommand{\record}{j_{\DG}}
\newcommand{\volDG}{\eps_{\DG}}
\newcommand{\psibar}{\overline{\psi}}
\newcommand{\sL}{S_{L}}
\newcommand{\sR}{S_{R}}

\title{Geometric selection of a single Weyl sector from symplectic orientation and a local causal record}
\author*[1]{\fnm{Alexey Sergeevich} \sur{Osipov}}\email{osipov.as@phystech.edu}
\author[2]{\fnm{Alexander Mettalinovich} \sur{Tishin}}
\affil[1]{\orgname{Moscow Institute of Physics and Technology}, \orgaddress{\street{9 Institutskiy Pereulok}, \city{Dolgoprudny}, \state{Moscow Region}, \postcode{141701}, \country{Russia}}}
\affil[2]{\orgname{Lomonosov Moscow State University}, \orgaddress{\street{1 Leninskie Gory}, \city{Moscow}, \postcode{119991}, \country{Russia}}}
\abstract{Why elementary fermionic degrees of freedom appear in inequivalent Lorentz-chiral representations is not answered by representation theory alone. We investigate a conditional geometric mechanism that selects one Weyl carrier without assigning electroweak gauge charges. Smooth unitary accessibility and regular characteristic reduction provide a four-dimensional symplectic quotient, whose orientation is stable under specified unitary and antiunitary observer transitions. An independently postulated local signed surface record is assumed to have a tensorial flux density and a leading, derivative-free, phase-invariant spinor bilinear in a restricted pre-Hodge Clifford-degree-three operator class. This bilinear has the established tensor structure of the variational Dirac spin current; the proposed identification with an operational record and its sign restriction remain physical hypotheses. Its Hodge dual is an axial flux. Under a nonzero leading coefficient and Lorentz-invariant complex-linear closure of elementary amplitudes, future-flux non-negativity selects exactly one Weyl subbundle. We also prove a convex-amplitude sufficient condition and give counterexamples showing why positivity alone does not select chirality. Parity-related realizations leave the absolute handedness label undetermined. We distinguish flux positivity from irreversible production, and describe independent implementation tests and failure criteria. This is the first structural step of a proposed research programme, not a derivation of Standard Model parity violation, gauge representations, or an experimentally verified record law.}
\begin{document}
\maketitle

\section{Introduction}\label{sec:introduction}
In four-dimensional Lorentzian spacetime the connected spin group admits two inequivalent irreducible two-component complex spinor modules, conventionally denoted $D(1/2,0)$ and $D(0,1/2)$. Both occur in the Dirac representation. The electroweak theory uses Lorentz chirality asymmetrically when assigning weak-isospin representations \cite{ref24,ref25}. Parity violation has independent experimental support \cite{ref8,ref9}. Lorentz representation theory alone, however, permits either chiral module. It does not specify a physical criterion that retains only one module within a proposed elementary carrier.

Theoretical attempts to explain or implement chiral asymmetry take substantially different starting points. Left--right symmetric models break a gauge symmetry dynamically \cite{ref26}; quantum cellular automata reconstruct relativistic free fields from information-theoretic assumptions \cite{darianoperinotti2017}; causal-fermion and spectral approaches encode fermionic structure in deeper geometric data \cite{finster2024,ref27}. Recent chiral-time geometric baryogenesis introduces dynamical CP-odd fields and cosmological evolution rather than selecting an elementary Lorentz module \cite{mir2026}. These precedents are important: the novelty claimed here is neither the use of geometry in fermion physics nor the idea of deriving physical laws from operational principles.

The dynamical use of chirality has several established forms. Chiral oscillations concern the evolution of existing Weyl components \cite{epjc_blasone2025}. Classification of Lorentz-covariant spinor fields is based on their bilinear structures \cite{epjc_lee2024}. Nontrivial spin topology can modify the relevant spinor dynamics \cite{epjc_dossantos2026}. Theory-space mass matrices may control chiral zero-mode profiles \cite{epjc_singh2026}, and Kaluza--Klein reductions can produce asymmetric chiral gauge couplings \cite{baptista2026}. Those mechanisms do not address the restricted kinematic question posed here. We ask whether independent geometric and causal conditions can select a single irreducible Lorentz-spin carrier before gauge quantum numbers have been assigned.

The language of operational distinguishability follows the broader programme of Refs.~\cite{ref1,ref2}. A recent independent bridge study by Tulupov and Tishin \cite{tulupov2026} identifies explicit closure-faithfulness and observer-identification conditions under which operational rank can pass to a smooth integral leaf and an observable Fisher rank. This does not prove a universal dimension-four theorem; in the present work the four-dimensional Lorentz sector and regular quotient remain declared inputs. The results below use only assumptions written in this article. In particular, the existence of a local irreversible current is not inferred from those preprints. The bridge to geometric quantum mechanics is motivated by the K\"ahler description of ray space \cite{ashtekarSchilling1997}. Observer covariance is implemented through the transition-probability version of Wigner's theorem \cite{bargmann1964}. These are established mathematical ingredients, with hypotheses that must still be verified for any proposed operational realization.

The physical issue is the meaning of a local signed record. Surface additivity and non-negativity are familiar characteristics of transported counting densities. A local differential-form representation, and the identification of its leading spinor part with a Clifford three-form, need stronger premises. We state them independently. In particular, the Dirac vector current is an admissible positive density if use of a spacetime orientation is allowed at the outset. That choice would not select a Weyl sector. The alternative selected here has the Clifford structure of the antisymmetric Dirac spin current. This provides a physical connection to the usual spin-response formalism \cite{hehl1976}, but no empirical identification of the postulated distinguishability record with that current is claimed.

\paragraph{Specific contribution and limits of priority.} The relevant question is whether \emph{a globally fixed relative orientation plus a local sign condition and an independently stated closure principle} can select an irreducible Lorentz carrier. The present contribution is a sufficiency theorem for that conjunction, together with counterexamples separating its essential hypotheses. We are not aware of a prior derivation that combines precisely these three ingredients into the same Weyl-carrier criterion. This limited literature-based statement does not claim that geometric chirality selection, operational field-theory reconstruction, or spinor-current duality is without precedent. Indeed, the relevant standard constructions are explicitly compared below.

Our statement is a theorem about a \emph{conditional} local carrier selection. It has three distinct layers: a four-dimensional geometry and spin lift, a signed local flux with a prescribed leading Clifford tensor class, and a closure condition on elementary amplitudes. The proof establishes the implication of their conjunction. We also give explicit counterexamples when the current or carrier restrictions are relaxed, thereby identifying where the selection mechanism can fail.

The resulting Weyl field is a formal building block. The construction supplies no weak-isospin or hypercharge assignments, no anomaly-free matter spectrum, no fermion masses, and no CP-violating observable. Its parity-related implementations yield opposite externally labelled chiralities. Sections~\ref{sec:setup}--\ref{sec:orientation} state the assumptions and construct the orientation. Section~\ref{sec:record} proves the flux and carrier statements. The remaining sections address parity, physical interpretation, effective corrections, quantum compatibility, and the logical status of the final theorem.

\section{Operational, geometric, and causal assumptions}\label{sec:setup}
\subsection{Physical spacetime and the domain of distinguishability}
Let $M$ be a smooth, connected, paracompact four-manifold without boundary, equipped with a Lorentzian metric $g$ of signature $(+---)$. A smooth unit timelike field $u$ defines the future cone:
\begin{equation}\label{eq:u}
 g(u,u)=1.
\end{equation}
At this stage no spacetime orientation or spin structure is specified. The four-dimensional Lorentz sector and the causal time orientation are empirical inputs. When making global spinor or quantum-field statements we impose global hyperbolicity (GH). The local algebraic theorem does not require GH.

A distinguishability-geometric (DG) domain is a smooth manifold $F_{\DG}$ equipped with a closed real two-form:
\begin{equation}\label{eq:Omega}
 \Omega\in\Omega^2(F_{\DG}),\qquad \D\Omega=0.
\end{equation}
The terminology does not embed field equations, an emergent metric theorem, or a dimension theorem. The operational principles C1--C5 discussed in Refs.~\cite{ref1,ref2} motivate distinguishability and a directed record. None of those principles is used as a hidden lemma.

\subsection{Reversible accessibility versus irreversible registration}
The mathematical distinction between accessing a state and registering an irreversible record is indispensable. In a quantum ray-space realization, smooth unitary evolution is Hamiltonian for the Fubini--Study symplectic form. For a self-adjoint generator $A$, the expectation-value function
\begin{equation}\label{eq:expHam}
 H_A([\psi])=\frac{\langle\psi,A\psi\rangle}{\langle\psi,\psi\rangle}
\end{equation}
generates the corresponding ray-space flow, up to the conventional normalization of the symplectic form \cite{ashtekarSchilling1997}. General measurement instruments, coarse graining, and dissipative semigroups are not Hamiltonian flows on the same symplectic space.

This restriction is visible already in Darboux coordinates. For $\Omega=\D x\wedge\D p$ and the contracting vector field $X=-\lambda(x\partial_x+p\partial_p)$,
\begin{equation}\label{eq:dissipative}
 \cL_X\Omega=-2\lambda\Omega\ne0\qquad(\lambda>0).
\end{equation}
Consequently, the Hamiltonian bridge below is a statement about smooth \emph{accessibility operations}. The irreversible registration of distinguishability is represented by a separate surface-record assumption. No claim is made that a general dissipative operation admits a Hamiltonian generator with respect to the same $\Omega$.

\paragraph{HCB (Hamiltonian continuum bridge).}
Every elementary accessible one-parameter procedure in the smooth reversible subcollection, when it has a differentiable generator, is represented locally by an admissible real observable $f$ and a Hamiltonian vector field $X_f$ satisfying
\begin{equation}\label{eq:hcb}
 \iota_{X_f}\Omega=\D f.
\end{equation}
Let $\cO$ be an accessibility orbit generated by these fields and their local compositional/Lie closure. Write
\begin{equation}\label{eq:orbit}
 T\cO=\LieSpan\{X_f:f\in\cA_{\mathrm{adm}}\}.
\end{equation}
If $\Omega$ is presymplectic, an $X_f$ exists only when $\D f$ annihilates the characteristic kernel. Admissible observables are restricted accordingly. Two representatives can differ by a kernel vector. This is a conditional smooth realization of accessible procedures, supported by geometric quantum mechanics in the unitary branch, and is not a theorem about all operational transformations.

\paragraph{RCR (regular characteristic reduction).}
Restrict $\Omega$ to an orbit $\cO$, set $K=\ker(\Omega|_{\cO})$, and assume that $K$ has locally constant rank. Its leaf space $P=\cO/K$ is assumed to be a smooth four-manifold with quotient submersion $q:\cO\to P$. The resulting reduced physical sector is identified by a specified smooth atlas with $M$. Thus
\begin{equation}\label{eq:rcr}
 \Omega_{\cO}=\iota_{\cO}^*\Omega,\qquad P=\cO/K,\qquad q^*\omega=\Omega_{\cO}.
\end{equation}
RCR includes the existence of a Hausdorff smooth quotient and the local identification with the Lorentz sector. A constant-rank condition by itself does not establish these global properties. Theorem~\ref{thm:reduction} derives nondegeneracy of $\omega$ once RCR holds. A composite preparation-and-readout scheme could operationally test rank by comparing tangent directions resolvable under enlarged families of controls; agreement for one restricted protocol is not sufficient evidence for a universal rank-four geometry \cite{tulupov2026}. This constraint prevents treating an assumed four-dimensional quotient as a derived prediction.

\subsection{Quantum observer covariance}
\paragraph{QOC (quantum observer covariance).}
For the observer-gluing step, the relevant charts $\cO_\alpha$ are realized as accessibility submanifolds of a projective Hilbert space. Allowed changes of observer are restrictions of \emph{bijective} ray-space transformations preserving every transition probability:
\begin{equation}\label{eq:dist}
 D_Q([\psi],[\phi])=1-\frac{|\langle\psi,\phi\rangle|^2}{\|\psi\|^2\|\phi\|^2}.
\end{equation}
The transitions must also map overlapping accessibility orbits to one another, obey the ordinary atlas cocycle on triple overlaps, and respect the specified quotient-to-$M$ identifications. Wigner's theorem then gives unitary or antiunitary implementers \cite{bargmann1964}, so for the restricted Fubini--Study forms
\begin{equation}\label{eq:qoc}
 \Phi_{\alpha\beta}^*\Omega_{\cO,\beta}=s_{\alpha\beta}\Omega_{\cO,\alpha},\qquad s_{\alpha\beta}\in\{+1,-1\}.
\end{equation}
Exact transition-probability preservation is stronger than merely agreeing on the rank ordering of distinguishability. For partial, non-surjective operational maps, the stated Wigner conclusion need not follow. QOC therefore specifies a physically meaningful, mathematically testable observer class. It is not inferred from generic operational consistency.

\subsection{A localized signed surface record}\label{subsec:edg}
The second bridge concerns records generated by irreversible registration. We assume a signed count transported through oriented spacelike hypersurface patches. Its physical interpretation is the net recorded throughput for a specified elementary spinor channel. In an implementation a detector must define what is counted and how the sign and normalization are calibrated. This empirical identification is part of a future realization of the framework.

In the representation-selection argument, $\psi$ denotes a commuting, complex-valued local spinor \emph{amplitude} or test section. The inequality on $N_\psi$ is an inequality of ordinary real numbers. An anticommuting Grassmann-valued classical fermion field has no canonical order relation and cannot directly satisfy such a positivity statement. Fermion parity here denotes the $\psi\mapsto-\psi$ grading of the amplitude expansion. Quantized anticommuting fields appear only after the elementary carrier is selected in Section~\ref{sec:qft}. This distinction is necessary for a mathematically meaningful positivity postulate.

\paragraph{EDG2$'$ (local causal surface record).}
Once a spin bundle is available, to each smooth local spinor configuration $\psi$ and compact oriented smooth spacelike three-chain $\Sigma$ assign a real number $N_\psi[\Sigma]$. We impose the following separate clauses.
\begin{enumerate}[label=(\roman*)]
\item \emph{Chain additivity.} The functional is linear on finite real combinations of oriented chains, including a reversal of sign under orientation reversal, and is local and continuous at order zero.
\item \emph{Tensorial flux locality.} There is a smooth real-valued alternating trilinear density $T_\psi(x;v_1,v_2,v_3)$ on tangent triples, linear in each argument and depending locally on the germ of the field at $x$ (including field jets when higher-derivative effective terms are considered). For every oriented spacelike chain,
\begin{equation}\label{eq:flux-density}
 N_\psi[\Sigma]=\int_\Sigma T_\psi(x;\cdot,\cdot,\cdot).
\end{equation}
The density contains no embedding-curvature or surface-jet dependence. Local absolute continuity with respect to the induced positive hypersurface measure is required. The tensorial clause is stronger than additivity over disconnected surface pieces.
\item \emph{Frame covariance and phases.} The record is covariant under proper orthochronous local spin-frame changes and invariant under $\psi\mapsto e^{i\alpha}\psi$. It is fermion parity even.
\item \emph{Minimal quadratic branch.} The first nonzero spinor-dependent term about $\psi=0$ is quadratic in $\psibar,\psi$ and contains no derivatives. At leading order the density is made from the Dirac pairing and Clifford multiplication in the \emph{pre-Hodge Clifford-degree-three class}: no $u$, reduced $\omega$, derived volume $\eps_{\DG}$, Hodge dual, Clifford volume element, or other externally supplied positive-rank tensor occurs in that term. This is a restricted physical operator hypothesis, not a consequence of Lorentz covariance alone. It excludes the orientation-dependent positive Dirac vector-current three-form; we therefore treat its justification as a separate, potentially falsifiable spin-source identification rather than concealing the choice in the word ``natural''.
\item \emph{Future-flux non-negativity.} On the designated elementary branch, the record of every sufficiently small patch with future unit normal is non-negative in the future orientation. The condition applies to every locally realizable admissible configuration.
\end{enumerate}

The first two clauses isolate a logical distinction. Additivity of an integral over surface pieces allows a density depending nonlinearly on a unit normal; it does not imply that the density is a differential form. The tensorial clause supplies the missing ingredient. Because the leading density contains no preferred future vector, imposing flux positivity before choosing the allowed elementary amplitudes would generally be inconsistent. The positivity clause is expressly restricted to the elementary branch specified next.

\paragraph{ELC (elementary linear-carrier closure).}
Let $S=\sL\oplus\sR$ be the complex Dirac spin bundle after a spin lift. The admissible elementary values at every $x$ form the fibre $E_x$ of a nonzero smooth complex linear subbundle $E\subseteq S$, invariant under the standard $\Spin^+(1,3)$ action in oriented local frames. Every $v\in E_x$ is realized by an admissible local smooth configuration, and sufficiently small real rescalings $t\psi$ remain admissible. Future-flux non-negativity applies to all such configurations.

ELC is motivated by the linear superposition principle for unnormalized elementary amplitudes in a fixed coherent representation sector. It does not follow from convexity of density operators, because density operators and spinor amplitudes have different linear structures. The carrier assumption also presupposes a coherent elementary sector: superselection rules or nonlinear selection rules may invalidate it. Section~\ref{subsec:weak-closure} gives a mathematically weaker, separately motivated alternative.

\subsection{The meaning of causal non-negativity}\label{subsec:flux-not-prod}
The sign condition in EDG2$'$ concerns flux through each sufficiently small future-oriented patch. It should be distinguished from non-negative \emph{production} of a cumulative quantity between two Cauchy surfaces. If $\record$ is a smooth three-form, Stokes' theorem gives
\begin{equation}\label{eq:balance}
 N_\psi[\Sigma_2]-N_\psi[\Sigma_1]=\int_V\D\record-\int_{B}\record,
\end{equation}
where $\partial V=\Sigma_2-\Sigma_1+B$ with the indicated boundary orientations. Even when the local surface density is positive, its total on successive surfaces may decrease. A separate balance equation $\D\record=\sigma\,\volDG$ with $\sigma\ge0$ (and specified lateral flux) would be needed to prove temporal production monotonicity. It is not assumed below, and no entropy-production theorem is claimed. The postulate under study is a positive elementary \emph{throughput} record whose underlying registration process may be irreversible.

\subsection{Two preliminary no-go observations}
\begin{proposition}[Dimension does not imply nondegeneracy]\label{prop:dimension}
Four-dimensionality and operational composition alone do not imply that a restricted closed two-form is symplectic.
\end{proposition}
\begin{proof}
On $\mathbb R^8$ let $\Omega=\sum_{a=1}^4\D q^a\wedge\D p_a$ and let the four-dimensional submanifold be $M=\{p_1=\cdots=p_4=0\}$. Its pullback of $\Omega$ vanishes. A Lorentz metric and a time orientation may be specified independently on $M$. Thus a reduction or nondegeneracy premise is needed.
\end{proof}

\begin{proposition}[A scalar arrow carries no chirality sign]\label{prop:arrow}
A scalar irreversible accumulation parameter together with a Lorentz metric and time orientation does not determine a sign convention for an axial spinor flux.
\end{proposition}
\begin{proof}
Given a scalar operational model, extend it once with the local spinor three-form $j$ and once with $-j$. Their scalar observables and causal cones agree. The two extensions give opposite flux signs. The distinction requires an additional spinor-valued record law or an independently calibrated signed observable.
\end{proof}

\section{Lorentz-spin representation data}\label{sec:spinors}
The connected spin double cover is $\Spin^+(1,3)\simeq SL(2,\mathbb C)$. Its complexified Lie algebra separates as
\begin{equation}\label{eq:spin-algebra}
 \mathfrak{spin}(1,3)\otimes\mathbb C\simeq\mathfrak{sl}_2(\mathbb C)_L\oplus\mathfrak{sl}_2(\mathbb C)_R.
\end{equation}
The inequivalent minimal modules are
\begin{equation}\label{eq:weyl-modules}
 \sL\simeq D(1/2,0),\qquad \sR\simeq D(0,1/2).
\end{equation}
The Dirac module is their direct sum. The labels are fixed only relative to a choice of oriented Clifford volume element. Reversing orientation interchanges the corresponding projectors. They are not weak-isospin labels. General background on the representation conventions can be found in Refs.~\cite{ref4,ref3,ref5,ref7,ref12}.

\section{Reduction, observer gluing, and spin structure}\label{sec:orientation}
\subsection{Symplectic reduction}
\begin{theorem}[Regular characteristic reduction]\label{thm:reduction}
Under HCB and RCR, there is a unique two-form $\omega$ on $P$ satisfying $q^*\omega=\Omega_{\cO}$. It is closed and nondegenerate.
\end{theorem}
\begin{proof}
For any local section $X$ of $K$, $\iota_X\Omega_{\cO}=0$ and
\begin{equation}\label{eq:basic}
 \cL_X\Omega_{\cO}=\D(\iota_X\Omega_{\cO})+\iota_X\D\Omega_{\cO}=0.
\end{equation}
The restricted form is horizontal and invariant along the characteristic leaves. It is therefore basic and descends to $P$ under the regular-quotient assumption \cite{ref10}. Since $q$ is a surjective submersion,
\begin{equation}\label{eq:closed-reduction}
 q^*(\D\omega)=\D(q^*\omega)=\D\Omega_{\cO}=0
\end{equation}
implies $\D\omega=0$. If $\omega(\bar v,\bar w)=0$ for every reduced vector $\bar w$, then a lift $v\in T\cO$ lies in $K$; hence $\bar v=0$. Thus $\omega$ is nondegenerate.
\end{proof}

\subsection{Unitary and antiunitary observer transitions}
\begin{lemma}[Descent through characteristic quotients]\label{lem:qoc-descent}
Suppose $\Phi_{\alpha\beta}:\cO_\alpha\to\cO_\beta$ satisfies Eq.~\eqref{eq:qoc}, with smooth inverse on the overlap. It maps $K_\alpha$ onto $K_\beta$ and induces a smooth quotient transition $\bar\Phi_{\alpha\beta}$ satisfying
\begin{equation}\label{eq:descend}
 \bar\Phi_{\alpha\beta}^*\omega_\beta=s_{\alpha\beta}\omega_\alpha.
\end{equation}
\end{lemma}
\begin{proof}
For $k\in K_\alpha$ and any $w\in T\cO_\alpha$,
\begin{equation}\label{eq:kernel-map}
 \Omega_{\cO,\beta}(\D\Phi\,k,\D\Phi\,w)=s_{\alpha\beta}\Omega_{\cO,\alpha}(k,w)=0.
\end{equation}
The inverse gives equality of kernels. Therefore $\Phi$ maps leaves to leaves and descends. Pulling back $q_\beta^*\omega_\beta=\Omega_{\cO,\beta}$, and using injectivity of pullback by the submersion $q_\alpha$, proves Eq.~\eqref{eq:descend}.
\end{proof}

\begin{theorem}[Observer-covariant global orientation]\label{thm:orientation}
If the reduced charts cover $M$ and satisfy RCR and QOC, the local symplectic squares define a globally compatible orientation of $M$.
\end{theorem}
\begin{proof}
On each reduced four-dimensional chart set
\begin{equation}\label{eq:symp-square}
 \nu_\alpha=\frac12\omega_\alpha\wedge\omega_\alpha.
\end{equation}
Since $\omega_\alpha$ is nondegenerate, $\nu_\alpha$ is nowhere zero. By Lemma~\ref{lem:qoc-descent},
\begin{equation}\label{eq:sign-square}
 \bar\Phi_{\alpha\beta}^*\nu_\beta=s_{\alpha\beta}^2\nu_\alpha=\nu_\alpha.
\end{equation}
Thus the local oriented-volume choices agree on overlaps. The quotient atlas and its cocycle condition identify them with a global orientation of $M$. Normalize it using $g$:
\begin{equation}\label{eq:normalize}
 \volDG=\frac{\nu_\alpha}{\lambda_\alpha},\qquad
 \lambda_\alpha=\left[-\frac1{4!}(\nu_\alpha)_{\mu\nu\rho\sigma}(\nu_\alpha)^{\mu\nu\rho\sigma}\right]^{1/2}.
\end{equation}
The sign is fixed by the symplectic square. The normalized forms agree under chart changes, giving a global unit volume form.
\end{proof}

In four real dimensions the sign of an antisymplectic observer change disappears upon squaring. This step uses more than a general orientation theorem for symplectic manifolds: the local reduced forms must be identified across actual observer charts. A failure of QOC or of the smooth quotient atlas can invalidate the construction.

\subsection{Global spin lift in the hyperbolic branch}
\begin{theorem}[Spin structure]\label{thm:spin}
A globally hyperbolic time-oriented four-dimensional Lorentz manifold endowed with the orientation of Theorem~\ref{thm:orientation} admits a global $\Spin^+(1,3)$ structure.
\end{theorem}
\begin{proof}
Global hyperbolicity gives a smooth splitting $M\simeq\mathbb R\times\Sigma$ with a Cauchy three-manifold $\Sigma$ \cite{ref13}. The restriction of $\iota_u\volDG$ orients $\Sigma$. Every orientable smooth three-manifold is parallelizable, so $T\Sigma$ is trivial. Consequently $TM$ is trivial and $w_2(TM)=0$. The obstruction to a spin structure vanishes \cite{ref14,ref11}.
\end{proof}
The spin structure need not be unique. Its isomorphism classes form a torsor under $H^1(M,\mathbb Z_2)$. The local algebraic selection is independent of that global choice, although global boundary conditions need not be.

\subsection{Global chirality operator}
Choose a spin structure and its Dirac bundle $S\to M$. With $\{\gamma^\mu,\gamma^\nu\}=2g^{\mu\nu}$, set
\begin{equation}\label{eq:gammafive}
 \gfive=\frac{i}{4!}\eps_{\DG,\mu\nu\rho\sigma}\gamma^\mu\gamma^\nu\gamma^\rho\gamma^\sigma.
\end{equation}
This expression is a global spin-equivariant endomorphism. In any positively oriented orthonormal frame it equals $i\gamma^0\gamma^1\gamma^2\gamma^3$ and satisfies
\begin{equation}\label{eq:gamma-properties}
 (\gfive)^2=1,\quad\{\gfive,\gamma(v)\}=0,\quad\nabla^S\gfive=0.
\end{equation}
It has globally defined eigensubbundles
\begin{equation}\label{eq:projectors}
\begin{aligned}
 P_L&=\frac{1-\gfive}{2},\qquad P_R=\frac{1+\gfive}{2},\\
 S&=\sL\oplus\sR.
\end{aligned}
\end{equation}
The orientation is essential to assigning the symbols $L$ and $R$ to the eigensubbundles. Their existence follows from the spin lift and does not yet select one of them as elementary.

\section{From the surface record to a Weyl carrier}\label{sec:record}
\subsection{A differential three-form and the exact role of tensoriality}
\begin{lemma}[Representation of a tensorial surface record]\label{lem:current}
Under EDG2$'$ clauses (i) and (ii), for each fixed smooth $\psi$ there is a unique smooth real three-form $\record(\psi)$ such that
\begin{equation}\label{eq:surface-record}
 N_\psi[\Sigma]=\int_\Sigma\record(\psi)
\end{equation}
for all admissible oriented spacelike three-chains $\Sigma$.
\end{lemma}
\begin{proof}
At each point $x$, the specified alternating trilinear map $T_\psi(x;\cdot,\cdot,\cdot)$ is, by definition, an element of $\Lambda^3T_x^*M$. Smoothness makes the family a smooth differential three-form $\record$. The surface integral reproduces the assumed local density, hence Eq.~\eqref{eq:surface-record}. If two smooth three-forms give the same integral on every sufficiently small oriented spacelike patch, their evaluations on each spacelike tangent three-plane agree. Those planes form an open set in $\Lambda^3T_xM$ and span it in dimension four, so the forms coincide pointwise.
\end{proof}

The lemma does not deduce tensoriality from chain additivity. A general order-zero local surface energy may depend on the direction of a unit normal through a nonlinear function and still be additive on separate patches. A flux instrument satisfies the additional tensorial clause if its infinitesimal signed response is linear in the oriented area three-vector, including under changes of patch tilt. This is an operational linear-response requirement that can be checked independently of the spinor sign test. In the language of currents, a smooth differential-form density is a specialization of the larger class of local functionals \cite{ref15}. Appendix~\ref{app:currents} gives the distinction explicitly. No conservation law $\D\record=0$ follows from Eq.~\eqref{eq:surface-record}. A deformation-invariance condition would be needed.

\subsection{Quadratic order and a restricted natural-map problem}
Fermion parity and phase invariance forbid odd powers of a classical commuting spinor amplitude in the smooth expansion about $\psi=0$. They do not force the first nonzero term to be quadratic: an admissible scalar functional could begin at quartic order. The minimal quadratic branch is therefore a separate clause of EDG2$'$. Since a nonzero natural three-form cannot be built from the metric alone without a further tensor, there is no leading field-independent term in the restricted pre-Hodge class. Writing $t\in\mathbb R$,
\begin{equation}\label{eq:quadratic}
 \record(t\psi)=t^2\record^{(2)}(\psi)+O(t^4),\qquad t\to0.
\end{equation}
The $O(t^4)$ remainder is local and smooth in the chosen admissible field family. The free Dirac field has engineering dimension $3/2$ in four dimensions, so a derivative-free bilinear three-form has canonical dimension three. The quadratic branch is the leading term of a low-field/low-energy operator expansion. The expansion alone does not determine its coefficient.

\subsection{Multiplicity and Clifford-degree-three minimality}
Fix $x\in M$ and put $V=T_xM$. The complexified Clifford algebra acts faithfully on the four-dimensional Dirac module and, as a spin representation,
\begin{equation}\label{eq:cliff-end}
 \End(S_x)\simeq\bigoplus_{r=0}^{4}\Lambda^rV_{\mathbb C}.
\end{equation}
The proper orthochronous equivariant bilinear maps from the Dirac pair to a complex three-form have multiplicity two. The corresponding representations satisfy
\begin{equation}\label{eq:hom-two}
 \dim_{\mathbb C}\Hom_{\Spin^+(1,3)}(\bar S\otimes S,\Lambda^3V_{\mathbb C}^*)=2.
\end{equation}
A convenient pair of representatives is
\begin{equation}\label{eq:two-threeforms}
 \psibar\gamma_{\mu\nu\rho}\psi,\qquad
 \psibar\gamma_{\mu\nu\rho}\gfive\psi.
\end{equation}
The second member explicitly uses the oriented Clifford volume element, or equivalently a Hodge operation. Hence Lorentz covariance and the field symmetries \emph{alone} leave two independent possibilities. An unrestricted uniqueness claim would be false.

\begin{lemma}[Restricted pre-Hodge natural-map uniqueness]\label{lem:unique}
Consider the real, phase-invariant, derivative-free quadratic three-form maps constructed by applying the Clifford-degree-three component $\gamma_{\mu\nu\rho}$ to the Dirac pairing, with no insertion of a volume element, Hodge map, preferred vector, $\omega$, or another external tensor. This specified class is one-dimensional over $\mathbb R$. Every nonzero member can be written
\begin{equation}\label{eq:cliff-three}
 \record^{(2)}(\psi)=\frac{\kappa(x)}{3!}\psibar\,i\gamma_{\mu\nu\rho}\psi\,\D x^\mu\wedge\D x^\nu\wedge\D x^\rho,
\end{equation}
with a real scalar coefficient $\kappa(x)$.
\end{lemma}
\begin{proof}
The grade-three summand in Eq.~\eqref{eq:cliff-end} gives one Clifford multiplication map to $\Lambda^3V_{\mathbb C}^*$. Inserting the grade-four volume element into that map gives the second representative in Eq.~\eqref{eq:two-threeforms}, which the class definition excludes. Any map in the specified grade-three class is therefore proportional to the first representative. In signature $(+---)$ the Dirac adjoint of $\gamma_{\mu\nu\rho}$ is minus itself. Multiplication by $i$ makes its Dirac bilinear real, leaving one real coefficient $\kappa(x)$.
\end{proof}

The lemma's scope is a physical operator selection rule. It is not a consequence of all naturality assumptions on Lorentz-spin bilinears. For example, once the spacetime orientation is available, the standard vector current
\begin{equation}\label{eq:vector-current}
 J^\mu=\psibar\gamma^\mu\psi
\end{equation}
can be converted into a three-form by $j_V=\iota_J\volDG$. Its flux is non-negative for \emph{all} Dirac spinors and selects neither chiral module. This is a direct counterexample to any claim that local covariance plus positivity, without the restricted pre-Hodge operator hypothesis, selects chirality.

\subsection{Independent physical motivation from the Dirac spin response}\label{subsec:spin-response}
A corresponding tensor appears in the established variational description of fermionic spin. Consider the Hermitian first-order Dirac Lagrangian on an orthonormal tetrad, temporarily treating the Lorentz spin connection $\omega_{\mu bc}$ as an independent source:
\begin{equation}\label{eq:dirac-lag}
 \cL_D=\frac{i}{2}\Big(\psibar\gamma^ae_a{}^\mu\nabla_\mu\psi-(\nabla_\mu\psibar)\gamma^ae_a{}^\mu\psi\Big)-m\psibar\psi.
\end{equation}
The spinorial connection acts through $\nabla_\mu\psi=\partial_\mu\psi+\frac14\omega_{\mu bc}\gamma^{bc}\psi$. Variation with respect to $\omega_{\mu bc}$ gives
\begin{equation}\label{eq:spin-current}
 \frac{\partial\cL_D}{\partial\omega_{abc}}
 =\frac{i}{8}\psibar\{\gamma^a,\gamma^{bc}\}\psi
 =\frac{i}{4}\psibar\gamma^{abc}\psi.
\end{equation}
Here $\gamma^{bc}=\gamma^{[b}\gamma^{c]}$, and $\omega_{abc}=e_a{}^\mu\omega_{\mu bc}$. The free Dirac spin response is totally antisymmetric and contains the same Clifford-grade-three bilinear of Eq.~\eqref{eq:cliff-three}. In four dimensions this bilinear is Hodge-dual to the axial current. The result is standard in spin-connection and torsion-based descriptions of fermionic matter \cite{hehl1976,shapiro2002}.

\begin{proposition}[A minimal spin-response route to the leading operator]\label{prop:spinroute}
Suppose that the spinor-dependent leading surface record is proportional to the fully antisymmetric response of the Hermitian, minimally coupled Dirac action to an independent Lorentz spin-connection source, with no higher-derivative source coupling at that order. Then its quadratic local three-form has the Clifford structure of Eq.~\eqref{eq:cliff-three}. This conclusion does not require a prior chirality projector or a spacetime volume orientation.
\end{proposition}
\begin{proof}
Minimal coupling gives the source derivative in Eq.~\eqref{eq:spin-current}. The identity $\{\gamma^a,\gamma^{bc}\}=2\gamma^{abc}$ shows that this response is fully antisymmetric, so its metric-lowered components define a local three-form bilinear. The Dirac adjoint makes $i\bar\psi\gamma_{abc}\psi$ real. Overall proportionality and local normalization are absorbed in a real $\kappa$.
\end{proof}

Equation~\eqref{eq:spin-current} provides an independent \emph{structural motivation} for the chosen pre-Hodge operator. Proposition~\ref{prop:spinroute} shows that the chosen tensor follows from a standard local source-coupling principle once the record is identified with the corresponding spin response. It does not demonstrate that an operational distinguishability record couples to the spin connection, measures spin density, or inherits positivity from the Dirac action. In conventional field theory the spin current is not generally a positive counting measure. The proposed physical bridge consists of the additional hypothesis that the local elementary record is carried by this spin-response channel and is subject to the sign condition in EDG2$'$. The source-coupling principle itself is familiar from field theory, while identifying the measured record with that source derivative is a separate hypothesis. Operationally, the source derivative measures the response to an infinitesimal local Lorentz-rotation coupling. This is a plausible channel for distinguishability between different spin preparations, because such preparations transform under the local spin group. A passive gauge change of frame alone is not a measurable perturbation, however. An actual coupling or spin-sensitive measurement protocol would have to be specified to give the response an operational interpretation. A realization must justify that identification by its measurement protocol or by a deeper operational dynamics.

This separation also clarifies the role of the metric. The tetrad and Dirac pairing enter the standard spin response. Neither $\eps_{\DG}$ nor a handedness projector is inserted in Eq.~\eqref{eq:spin-current}. The discrete sign enters only when the three-form is dualized using the orientation obtained in Section~\ref{sec:orientation}.

\subsection{Hodge duality and sign conventions}\label{subsec:dual}
For the normalized oriented volume form $\volDG$ there is a unique vector field $\Jax$ satisfying
\begin{equation}\label{eq:dual}
 \record^{(2)}=\iota_{\Jax}\volDG.
\end{equation}
Our convention $\eps_{0123}=+1$, $\gamma^5=i\gamma^0\gamma^1\gamma^2\gamma^3$ in a positively oriented orthonormal frame yields the Clifford identity
\begin{equation}\label{eq:cliff-identity}
 i\gamma_{\mu\nu\rho}=+\eps_{\DG,\mu\nu\rho\sigma}\gamma^\sigma\gfive.
\end{equation}
The sign in Eq.~\eqref{eq:cliff-identity} is fixed by these definitions. It is not a free normalization choice. Inserting it in Eqs.~\eqref{eq:cliff-three} and \eqref{eq:dual} gives
\begin{equation}\label{eq:axial}
 \Xi_{\DG}^{\mu}=-\kappa\psibar\gamma^\mu\gfive\psi.
\end{equation}
For example, with $\gamma_{123}=-\gamma^1\gamma^2\gamma^3$ and $\eps_{1230}=-1$, Eq.~\eqref{eq:cliff-identity} gives $i\gamma_{123}=-\gamma^0\gamma^5$, so $\Xi^0_{\DG}=-\kappa\psibar\gamma^0\gamma^5\psi$, as Eq.~\eqref{eq:axial} requires. Reversing the DG orientation while keeping the primary three-form fixed reverses $\Xi_{\DG}$.

\subsection{From integral non-negativity to a pointwise inequality}
For a spacelike patch $U$ with future unit normal $n$, choose its orientation so that the induced positive measure $\D\Sigma$ is the pullback of $\iota_n\volDG$. Then
\begin{equation}\label{eq:fluxlocal}
 \record^{(2)}\big|_U=(n_\mu\Xi^\mu_{\DG})\,\D\Sigma.
\end{equation}
\begin{lemma}[Localization of flux positivity]\label{lem:pointwise}
For a smooth local record, non-negativity of the integral on every sufficiently small future-oriented spacelike patch is equivalent to non-negativity of its pointwise future-normal density.
\end{lemma}
\begin{proof}
If $n_\mu\Xi^\mu_{\DG}<0$ at a point of such a patch, continuity makes the density negative on a smaller neighbourhood, contradicting its integral non-negativity. Conversely, the integral of a non-negative density is non-negative.
\end{proof}

For a future unit timelike $n$ the Hermitian form
\begin{equation}\label{eq:positive-pairing}
 (\varphi,\psi)_n=\bar\varphi\gamma(n)\psi
\end{equation}
is positive definite. The eigensubbundles of $\gfive$ are orthogonal for this pairing, because $\gfive$ is self-adjoint with respect to it. Writing $\psi=\psi_L+\psi_R$,
\begin{equation}\label{eq:rho}
 \rho_{\DG}^{(2)}(\psi;n)=n_\mu\Xi_{\DG}^{\mu}
 =\kappa\big(\|\psi_L\|_n^2-\|\psi_R\|_n^2\big).
\end{equation}

\subsection{Selection theorem with an arbitrary nonzero record sign}
\begin{theorem}[Conditional single-Weyl-carrier selection]\label{thm:selection}
Assume EDG2$'$, ELC and a smooth \emph{nowhere-zero} coefficient $\kappa$ for the quadratic record on connected $M$. Then the admissible elementary carrier is a single irreducible Weyl subbundle. More specifically,
\begin{equation}\label{eq:choice}
 E=\begin{cases}\sL,&\kappa>0,\\\sR,&\kappa<0.\end{cases}
\end{equation}
The sign of $\kappa$ is constant on connected $M$.
\end{theorem}
\begin{proof}
A continuous nowhere-zero real function on connected $M$ has constant sign. At each $x$, the finite-dimensional Dirac module is the direct sum of two inequivalent irreducible complex $\Spin^+(1,3)$ modules. Its invariant complex linear subspaces are precisely $0$, $(\sL)_x$, $(\sR)_x$ and $(\sL\oplus\sR)_x$.

By ELC, every $v\in E_x$ is the value at $x$ of an admissible local field $\psi$ and $t\psi$ remains admissible for sufficiently small real $t$. Equations~\eqref{eq:quadratic}, \eqref{eq:rho} and Lemma~\ref{lem:pointwise} give
\begin{equation}\label{eq:limit-positivity}
 0\le\lim_{t\to0}\frac{\rho_{\DG}(t\psi;n)}{t^2}
 =\kappa(x)\big(\|v_L\|_n^2-\|v_R\|_n^2\big).
\end{equation}
If $\kappa>0$, neither $(\sR)_x$ nor the full Dirac fibre obeys Eq.~\eqref{eq:limit-positivity}, whereas $(\sL)_x$ does. The nonzero condition excludes the zero subspace, leaving $E_x=(\sL)_x$. If $\kappa<0$, the roles are reversed. Smoothness of $E$ then gives Eq.~\eqref{eq:choice} globally.
\end{proof}

The nonzero-coefficient condition is needed at every point. If $\kappa$ vanishes, higher-order terms may control the sign there, and the leading-order theorem alone does not decide the fibre. The original positive-calibration case is recovered by taking $\kappa>0$ throughout. Eq.~\eqref{eq:choice} makes explicit which sign premise is relevant.

\subsection{A mixed Dirac counterexample without carrier closure}\label{subsec:mixed-example}
Positive future flux alone does not force a pure Weyl spinor. Consider Minkowski space in the Weyl representation, where $\gamma^5=\operatorname{diag}(-I_2,I_2)$. Choose
\begin{equation}\label{eq:mixed-spinor}
 \psi_L=\begin{pmatrix}1\\0\end{pmatrix},\qquad
 \psi_R=\begin{pmatrix}0\\a\end{pmatrix},\qquad 0<|a|<1.
\end{equation}
Both chiral components are nonzero. Their vector currents are
\begin{equation}\label{eq:two-null}
 J_L^\mu=(1,0,0,-1),\qquad
 J_R^\mu=|a|^2(1,0,0,-1).
\end{equation}
For $\kappa>0$, Eq.~\eqref{eq:axial} gives
\begin{equation}\label{eq:mixed-current}
 \Xi_{\DG}^\mu=\kappa(1-|a|^2)(1,0,0,-1).
\end{equation}
This vector is future null, so $n_\mu\Xi_{\DG}^{\mu}>0$ for every future unit timelike $n$. Proper orthochronous Lorentz transformations preserve its future causal character. Thus there exist Lorentz-covariant sets of mixed Dirac amplitudes satisfying causal non-negativity. Such sets need not be complex-linear. ELC excludes them by requiring closure of the entire invariant carrier under coherent linear combinations. The example proves that the closure premise carries independent logical content.

\subsection{A convex-amplitude alternative to ELC}\label{subsec:weak-closure}
The full vector-space assumption can be replaced by a geometric condition on accessible unnormalized amplitudes. This observation does not remove the need for an operational amplitude-closure principle.
\begin{proposition}[Convex phase-closed carrier alternative]\label{prop:convex}
At each $x$ let $\cA_x\subset S_x$ be a nonempty set containing a nonzero amplitude. Suppose it is convex over real convex combinations, invariant under multiplication by every complex phase, invariant under $\Spin^+(1,3)$, and all its values are locally realizable by configurations to which the smooth quadratic flux inequality applies, including sufficiently small amplitudes. If $\kappa(x)\ne0$, then $\cA_x$ equals the selected Weyl fibre appearing in Eq.~\eqref{eq:choice}.
\end{proposition}
\begin{proof}
Phase invariance supplies $-v$ with every $v\in\cA_x$, and convexity then gives $0\in\cA_x$. Let $E_x=\Span_{\mathbb C}(\cA_x)$. For $w=\sum_{j=1}^m c_jv_j$ with $v_j\in\cA_x$, choose $t>0$ so that $t\sum_j|c_j|\le1$. The amplitude $tw$ is a convex combination of $0$ and $e^{i\arg c_j}v_j$, with nonnegative weights $1-t\sum_j|c_j|$ and $t|c_j|$. Hence $tw\in\cA_x$. Divide the quadratic flux inequality for $tw$ by $t^2$ and send $t\to0$. It follows that the quadratic form in Eq.~\eqref{eq:limit-positivity} is non-negative for every $w\in E_x$. Since $E_x$ is a nonzero invariant complex subspace, the representation argument in Theorem~\ref{thm:selection} gives $E_x=(\sL)_x$ or $(\sR)_x$, according to $\operatorname{sgn}\kappa$.

The natural $SL(2,\mathbb C)$ action on either two-component Weyl module is transitive on its nonzero vectors. As $\cA_x$ is invariant and contains a nonzero vector of the selected module, it contains every nonzero vector of that module, together with zero. Thus $\cA_x=E_x$.
\end{proof}

Convexity is appropriate for an unnormalized \emph{amplitude preparation set} only if coherent interpolation of accessible preparations is physically allowed. The convexity of mixed-state density matrices is insufficient for this proposition, and the condition must be justified in a particular operational realization. It provides an alternative route to the same conclusion with a different, explicitly identified physical assumption.

\section{Parity underdetermination and smooth branches}\label{sec:parity}
\subsection{Two parity-related models}
Fix an external coordinate orientation $\eps_0=\D t\wedge\D x\wedge\D y\wedge\D z$ on $\mathbb R^{1,3}$, with $g=\D t^2-\D x^2-\D y^2-\D z^2$ and $u=\partial_t$. Define
\begin{equation}\label{eq:two-omegas}
 \Omega_A=\D t\wedge\D x+\D y\wedge\D z,\qquad
 \Omega_B=\D t\wedge\D x-\D y\wedge\D z.
\end{equation}
They are closed, symplectic and satisfy
\begin{equation}\label{eq:omegas-squared}
 \tfrac12\Omega_A^2=+\eps_0,\qquad
 \tfrac12\Omega_B^2=-\eps_0.
\end{equation}
Take each DG domain and reduced physical sector to be the whole Minkowski manifold, let the admissible smooth compactly supported Hamiltonians generate the accessibility orbit, and set $K=0$. HCB and RCR are satisfied. Each model has a one-chart QOC description. Appendix~\ref{app:models} supplies separate quantum-ray-space examples with nonidentity observer transitions; these should not be confused with a global symplectomorphism between the constant flat form and a product Fubini--Study form.

Use the same pre-Hodge law (with one fixed nonzero $\kappa$) in both models. Their orientation-dependent chirality operators satisfy
\begin{equation}\label{eq:opposite-gamma}
 \gamma_B^5=-\gamma_A^5.
\end{equation}
Theorem~\ref{thm:selection} selects the negative eigensubbundle of $\gamma_A^5$ and $\gamma_B^5$ when $\kappa>0$, respectively. For $\kappa<0$ it selects the positive eigensubbundles. In either case the two models choose opposite chirality relative to the \emph{same} external reference orientation.

\begin{theorem}[Parity underdetermination]\label{thm:parity}
The common geometric and causal assumptions do not determine the name of the selected Weyl sector relative to an externally fixed spatial orientation. They determine the existence of exactly one selected invariant Weyl carrier under the hypotheses of Theorem~\ref{thm:selection}.
\end{theorem}
\begin{proof}
The models in Eq.~\eqref{eq:two-omegas} have opposite DG orientations relative to $\eps_0$, while sharing the metric, future cone, pre-Hodge Clifford law and specified regularity assumptions. Theorem~\ref{thm:selection} applies to each, and Eq.~\eqref{eq:opposite-gamma} exchanges their externally labelled selected modules. Thus an absolute left/right assertion cannot follow from the common hypotheses, whereas single-sector selection does.
\end{proof}

This is model-theoretic underdetermination. It does not derive spontaneous parity breaking. A laboratory convention can label the realized branch, but that convention adds information not present in the parity-covariant assumptions.

\subsection{No smooth flip within one regular connected branch}
\begin{proposition}[Smooth-domain obstruction]\label{prop:wall}
Suppose $M$ is connected, the reduced symplectic orientation is globally smooth and nowhere zero, and $\kappa$ is smooth and nowhere zero. The selected externally labelled Weyl sector cannot change continuously across an internal hypersurface while these assumptions remain true.
\end{proposition}
\begin{proof}
Relative to any fixed continuous reference orientation, the ratio of two nonvanishing four-forms has a continuous sign and therefore constant sign on connected $M$. The sign of $\kappa$ is likewise constant. Equation~\eqref{eq:choice} then gives the same externally labelled choice throughout $M$.
\end{proof}
The proposition allows interfaces where the reduced two-form becomes degenerate, $\kappa$ vanishes, the orientation fails to glue, or the regularity premises break down. It gives no cosmological initial-condition theory.

\section{Physical content, independent assumptions, and possible tests}\label{sec:interpretation}
\subsection{Which steps follow from existing theory?}
The assumptions admit a layered physical interpretation. For closed systems with unitary dynamics, the Hamiltonian description of ray-space accessibility is standard geometric quantum mechanics \cite{ashtekarSchilling1997}. Regular reduction is a standard presymplectic construction, conditional on the smooth quotient. If observer changes are bijective and preserve transition probabilities, the unitary/antiunitary dichotomy follows from Wigner's theorem \cite{bargmann1964}. These statements support HCB and QOC in restricted domains, but do not establish their validity for every irreversible operational procedure.

The principal new physical hypothesis concerns EDG2$'$. Its first part states that irreversible registration can be localized as an additive signed current on spacelike surface patches. Its tensorial clause is the operational content needed for the representation by a three-form. Its Clifford restriction identifies the leading spinor signal with the antisymmetric spin response found in conventional Dirac field theory. Its positivity clause chooses an elementary counting convention for this response. Each of these premises can in principle fail in a physical realization. For example, nonlinear normal dependence would invalidate Lemma~\ref{lem:current}; an orientation-dependent vector-current record would invalidate Lemma~\ref{lem:unique}; an accessible mixed amplitude set without coherent closure would invalidate Theorem~\ref{thm:selection}.

ELC introduces a separate quantum-linear assertion about a single elementary carrier. The necessity of some closure principle follows from the explicit mixed-spinor model in Section~\ref{subsec:mixed-example}. The theorem in Section~\ref{subsec:weak-closure} identifies one alternative closure formulation. Neither is obtained solely from the geometric axioms for distinguishability.

\subsection{An independent-response test for the spin-record hypothesis}\label{subsec:spin-test}
The structural identity of the Clifford three-form with a spin-connection response is not an identification of measurable observables. A non-circular physical realization should specify a \emph{source protocol} before the chirality test: a controlled infinitesimal coupling to the local Lorentz-spin generator, a corresponding differential response between calibrated spin preparations, and an oriented surface procedure measuring the putative signed throughput. The coefficient $\kappa$, including its sign, must be fixed independently of the observed selected module. Only then can the inequality in EDG2$'$ test a new physical statement rather than restate a choice of carrier.

The distinguishing diagnostic is the response to source and detector geometry, not an arbitrary relabelling of spinor components. On sufficiently small patches one should compare the measured dependence on the area three-vector with a linear tensorial current model. A non-linear dependence on the unit normal would falsify the three-form postulate even if all flux readings were non-negative. One should separately compare source response to the axial candidate with the conventional positive Dirac vector-number current. A measurement governed only by the latter fails the proposed selection criterion rather than validating it. Finally, tests of closure must concern \emph{coherently prepared amplitudes} rather than mixtures of density operators.

A passive rotation of the spin frame is a gauge relabelling and cannot itself be counted as an active measurement. The source protocol requires an independently controlled physical perturbation, for example an effective spin-connection/torsion coupling in a suitable analogue or gravitational response model. No such experimentally realized source detector is claimed in this paper. Specifying this sequence makes clear what would have to be measured to assess EDG2$'$ without fitting the assumptions retrospectively.

\subsection{Operational interpretation of the signed flux}
A realization of EDG2$'$ must specify how a signed count on an oriented detector patch is registered. The choice of the detector normal supplies the positive hypersurface measure; reversing the patch orientation reverses the signed record. The internal sign of the leading Dirac spin-response contribution is measured relative to that orientation. For a proposed elementary branch, the physically asserted property is non-negativity of its calibrated future-flux density at sufficiently small amplitude. Such a statement is stronger than choosing a positive sign for an arbitrary mathematical bilinear, because the sign and normalization must be independently determined by the measurement convention before using the selection theorem.

The free massless Weyl equation illustrates the distinction between registration and production. On the selected sector Eq.~\eqref{eq:axial} is proportional to the usual vector current. For a free on-shell Weyl field this current is conserved, so $\D\record=0$ and, in the absence of boundary leakage, its total flux is constant across Cauchy surfaces. An irreversible detector can register a conserved flux without the flux itself being an entropy-production current. Equation~\eqref{eq:balance} is the appropriate balance law if additional production processes are introduced. Claims about a strictly increasing count require a separately specified irreversible dynamics or source term.

\subsection{Discriminating consequences and their limits}
A concrete physical implementation would require the following independent inputs. The accessibility form and its reduced orientation should be reconstructible from transition-probability data or another calibrated quantum-geometric protocol. The record should be measurable on small spacelike patches and satisfy the tensorial additivity test, including its dependence on patch tilt. The leading field dependence should be quadratic, with a measured spin-response rather than a vector-number-current structure. A coherent preparation protocol should determine whether the accessible elementary amplitudes satisfy ELC or the convex-amplitude alternative. The sign test on those amplitudes would then be a nontrivial consequence of the independent calibrations.

The present manuscript does not supply such a protocol or a dynamical model that enforces EDG2$'$. The proposed selection is therefore a physically motivated \emph{conditional criterion}, with explicit failure modes. In particular, observing a purely positive Dirac vector density is no evidence for the axial-record hypothesis. A test of gauge-sector parity violation would additionally require a map from the selected Lorentz block to internal representations and interactions. The present theorem does not predict the empirical weak-isospin assignments of the Standard Model.

\subsection{Comparison with nearby mechanisms}\label{subsec:comparison}
The distinction between selection of a Lorentz representation and dynamical chiral asymmetry is material. Standard weak-interaction model building assigns different gauge representations to left- and right-handed fields. Left--right theories introduce a parity-related gauge completion and then break it \cite{ref26}. Theory-space mechanisms engineer massless chiral zero modes by algebraic constraints on mass matrices \cite{epjc_singh2026}. Kaluza--Klein scenarios obtain chiral couplings from higher-dimensional reduction \cite{baptista2026}. Chiral-time equivalence links geometric pseudoscalars to cosmological baryon-number production \cite{mir2026}. Operational quantum cellular automata derive free-field kinematics from discrete interaction assumptions \cite{darianoperinotti2017}, whereas causal fermion systems and noncommutative geometry investigate broader spacetime and internal-algebraic constructions \cite{finster2024,ref27}. None of these claims that a generic positive axial current alone selects a Weyl module; neither do we.

Our result has a narrower mathematical target: a physically motivated \emph{candidate signed record} plus a geometric orientation and an explicit elementary-carrier closure imply a single Lorentz-Weyl fibre. The new statement is the combined sufficiency/obstruction result, not the underlying Clifford identity, spin-current variation, symplectic reduction, or classification of irreducible spin representations. The mixed Dirac counterexample is indispensable because it proves that sign positivity on individual spinors, even for all future normals, is not enough. The Dirac-vector-current counterexample is equally indispensable because it shows that covariance and positivity do not isolate the desired current without independent operator input.

\subsection{Novelty relative to standard currents and geometric mechanisms}
The symplectic reduction, global spin structure, Dirac spin-current identity and algebra of Weyl projectors are existing mathematical and physical results. Their roles are isolated in Sections~\ref{sec:orientation} and \ref{sec:record}. The proposed extension is the combination of an observer-covariant DG orientation with a restricted, signed, local spin-response record and an elementary-carrier closure condition, resulting in the conditional implication of Theorem~\ref{thm:selection}. This differs in logical target from chiral zero-mode localization \cite{epjc_singh2026}, geometric modifications of Dirac dynamics \cite{epjc_dossantos2026}, and Kaluza--Klein chiral couplings \cite{baptista2026}.

A possible objection is that postulating a positive axial current already amounts to selecting chirality. Section~\ref{subsec:mixed-example} shows that this is not true for individual amplitudes: mixed Dirac spinors can have positive axial flux for every future observer. The restriction to one module follows only when the admissible amplitudes possess the additional closure property. Conversely, Section~\ref{subsec:dual} demonstrates that the axial current is selected only after the pre-Hodge operator-class assumption is stated. The paper's foundational question is therefore whether those two additional operational features admit independent physical realizations, not whether chirality follows from Lorentz invariance alone.

\section{Effective corrections and the regime of stability}\label{sec:eft}
\subsection{Scope of the effective description}
The theorem concerns the leading derivative-free Clifford three-form. A more general effective record may depend on derivatives, curvature, gauge fields or DG tensors. Its operator content depends on what additional structures are permitted. We consider a restricted test basis to make no claim of a complete effective-field-theory classification.

Assume a torsion-free spin connection with $\nabla\gfive=0$, no chiral mass mixing in the selected massless carrier, and no external parity-odd tensor or explicit insertion of $u$ into the building block. The vector $u$ is used only to evaluate a future-normal flux. Terms outside this stated operator class require their own analysis.

\subsection{An algebraic vanishing identity}
\begin{lemma}[Same-chirality even-Clifford bilinears]\label{lem:even}
For a pure Weyl section $\psi_L=P_L\psi$ and any Clifford-even element $\Gamma_{\mathrm{even}}$,
\begin{equation}\label{eq:even}
 \bar\psi_L\Gamma_{\mathrm{even}}\nabla_{\mu_1}\cdots\nabla_{\mu_r}\psi_L=0
\end{equation}
for every $r\ge0$, and likewise for $R$.
\end{lemma}
\begin{proof}
The Dirac adjoint exchanges the chirality projectors, so $\bar\psi_L=\bar\psi P_R$. Even Clifford elements commute with $\gfive$ and its projectors. Since $\nabla P_L=0$,
\begin{equation}\label{eq:even-proof}
 \bar\psi_L\Gamma_{\mathrm{even}}\nabla^r\psi_L
 =\bar\psi P_R\Gamma_{\mathrm{even}}P_L\nabla^r\psi=0.
\end{equation}
\end{proof}

Define the illustrative one-derivative real span
\begin{equation}\label{eq:derivatives}
 \begin{split}
 \cB_{1\partial}&=\Span_{\mathbb R}\{\cO_1,\cO_2\},\\
 \cO_1&=\D(\psibar\gamma_{(2)}\psi),\\
 \cO_2&=\Alt(\psibar\gamma_{(2)}\overleftrightarrow{\nabla}\psi).
 \end{split}
\end{equation}
Here $\gamma_{(2)}$ is the Clifford-valued two-form, and $\Alt$ antisymmetrizes the derivative and Clifford indices. The bidirectional derivative is $\psibar\gamma_{(2)}\nabla\psi-(\nabla\psibar)\gamma_{(2)}\psi$. Both candidates vanish on a pure Weyl section by Lemma~\ref{lem:even}. The displayed pair is a \emph{test basis}. It is not asserted to exhaust all dimension-four or one-derivative three-form operators.

\subsection{Uniform-smallness theorem}
Let the future density on normalized admissible jet configurations be
\begin{equation}\label{eq:eff-rho}
 \rho_{\mathrm{eff}}=\rho_0+\delta\rho,\qquad \|\psi\|_u=1,
\end{equation}
where $\rho_0$ is Eq.~\eqref{eq:rho}. In the following theorem assume a fixed-sign coefficient satisfying $|\kappa(x)|\ge\kappa_*>0$ on the regime considered.

\begin{theorem}[Stability under uniformly small corrections]\label{thm:stability}
Suppose ELC holds for the effective carrier and for every normalized jet in the domain of the effective description
\begin{equation}\label{eq:small-bound}
 |\delta\rho(\psi;x)|\le\eta\,|\kappa(x)|,\qquad 0\le\eta<1.
\end{equation}
Then the positive and negative signs of the two pure Weyl sectors remain separated, and the unique maximal nonzero invariant elementary carrier selected by the effective inequality is the same as in Theorem~\ref{thm:selection}.
\end{theorem}
\begin{proof}
On the sector with positive leading density, $\rho_{\mathrm{eff}}\ge(1-\eta)|\kappa|>0$. On the other sector $\rho_{\mathrm{eff}}\le-(1-\eta)|\kappa|<0$. The invariant-subspace classification used in Theorem~\ref{thm:selection} then excludes the latter sector and the full Dirac module. The selected sector remains admissible throughout the stipulated normalized domain.
\end{proof}

The bound in Eq.~\eqref{eq:small-bound} is a strong \emph{uniform} hypothesis; it does not follow from canonical dimension counting alone. Derivative corrections can become large for high-frequency fields, and curvature or background insertions can introduce further contributions. The theorem is a sign-continuity statement over a controlled effective regime. It does not establish all-scale protection or an operator-classification no-go.

\section{Locally covariant quantum-field compatibility}\label{sec:qft}
\subsection{CAR algebra and one-particle setting}
On a globally hyperbolic spin manifold the free massless Dirac and Weyl systems have well-posed Cauchy problems. The Weyl subbundle selected by Theorem~\ref{thm:selection} supports a locally covariant CAR algebra. This construction requires neither a preferred vacuum nor a globally defined positive-frequency splitting \cite{ref16,ref17,ref18}. A Hadamard state yields a representation in which suitably renormalized local composite fields can be defined \cite{ref19,ref20}.

In stationary Minkowski space a selected left-handed field can be expanded schematically as
\begin{equation}\label{eq:fock-field}
 \hat\psi_L(x)=\int\D\Pi_p\,[a(p)u_L(p)e^{-ipx}+b^\dagger(p)v_L(p)e^{ipx}].
\end{equation}
The antiparticle created by $b^\dagger$ belongs to the standard quantization of the selected Weyl field \cite{ref6}. It is not an additional independent elementary right-Weyl field. The analogous construction applies when $\kappa<0$ and the selected elementary bundle is $S_R$.

\subsection{Locality and one-particle flux}
The quadratic signed current is represented by a renormalized Wick product,
\begin{equation}\label{eq:wick}
 \hat\Xi_{\DG}^\mu=-\kappa:\!\hat{\bar\psi}\gamma^\mu\gfive\hat\psi\!:\ ,
\end{equation}
with the standard state/geometry-dependent renormalization qualifications on curved backgrounds. Its smeared bilinear observables commute at spacelike separation: exchanges of even products of fermionic fields produce no graded sign. This is compatible with local covariance of the CAR algebra.

In a stationary flat region and for a non-negative compactly supported function $h$ on a Cauchy surface, define
\begin{equation}\label{eq:smeared-charge}
 \hat N_{\DG}[h]=\int_\Sigma h\,n_\mu\hat\Xi_{\DG}^\mu\,\D\Sigma.
\end{equation}
On the selected elementary one-particle branch,
\begin{equation}\label{eq:particle-positive}
 \langle f|\hat N_{\DG}[h]|f\rangle
 =|\kappa|\int_\Sigma h(x)|f(x)|^2\,\D\Sigma\ge0.
\end{equation}
This is a statement about a specified stationary particle representation and normally ordered density, not a state-independent pointwise positivity theorem for arbitrary Hadamard states.

\subsection{Antiparticle sign and full-Fock obstruction}
For a one-antiparticle state the same signed charge satisfies
\begin{equation}\label{eq:anti-negative}
 \langle\bar g|\hat N_{\DG}[h]|\bar g\rangle
 =-|\kappa|\int_\Sigma h(x)|g(x)|^2\,\D\Sigma\le0.
\end{equation}
For the unsmeared total operator in a stationary flat Fock representation,
\begin{equation}\label{eq:fock-charge}
 \hat Q_{\DG}=|\kappa|\int\D\Pi_p\,[a^\dagger(p)a(p)-b^\dagger(p)b(p)].
\end{equation}
\begin{theorem}[No full-Fock positivity]\label{thm:fock}
The signed operator $\hat Q_{\DG}$ is not positive on the full stationary Fock space; its spectrum is unbounded below.
\end{theorem}
\begin{proof}
An $m$-antiparticle state has eigenvalue $-m|\kappa|$. Arbitrarily large finite $m$ produce an unbounded-below spectrum. No finite scalar renormalization shift can make the operator positive.
\end{proof}

The positivity postulate is imposed on the \emph{elementary classical carrier} before second quantization. It is consistent with the existence of antiparticles because quantized currents behave as signed charges. Replacing $a^\dagger a-b^\dagger b$ by a particle-plus-antiparticle number requires a positive-frequency splitting, which is generally state dependent and nonlocal on curved spacetimes. Hence a universal positive particle-number current would be a different assumption.

\subsection{Anomalies and the scope of CPT}
In stationary Minkowski space, the usual CPT operation interchanges particles and antiparticles and reverses the corresponding signed charge. The existence of two parity-related orientation branches does not by itself imply fundamental CPT breaking. A general curved spacetime need not carry a global discrete CPT isometry. Chiral gauge anomalies, trace anomalies and curvature-induced fermion effects remain separate dynamical questions \cite{larue2023,kamada2025,maleknejad2025}. A gauge-coupled completion must satisfy the appropriate anomaly-cancellation conditions. The current theorem supplies no such completion.

\section{Main conditional theorem and assumption map}\label{sec:main}
\begin{theorem}[Weyl-sector selection under geometric and causal inputs]\label{thm:main}
Assume a connected four-dimensional time-oriented Lorentz manifold $(M,g,u)$; a smooth DG domain with closed real two-form; HCB and RCR with a smooth four-dimensional quotient identified with $M$; QOC for the observer overlaps in a ray-space realization; and EDG2$'$ on a smooth Dirac bundle with a nonzero leading quadratic coefficient $\kappa$. Assume ELC, or the alternative hypotheses of Proposition~\ref{prop:convex}. Use GH for the global spinor and CAR statements. Then the following hold.
\begin{enumerate}[label=(\arabic*)]
\item The reduced two-form is symplectic and its square defines a global orientation of $M$.
\item In the globally hyperbolic branch the oriented manifold admits a global spin structure and a chirality splitting $S=\sL\oplus\sR$.
\item The tensorial surface record has a unique differential three-form density, and its leading term is the Clifford-grade-three bilinear within the specified restricted pre-Hodge class.
\item Hodge duality makes this leading record an axial current with the definite sign of Eq.~\eqref{eq:axial}.
\item Future causal non-negativity and the carrier-closure condition select exactly one Weyl subbundle according to Eq.~\eqref{eq:choice}.
\item The external absolute label of that subbundle is parity-underdetermined; a smooth sign-changing domain within one regular connected branch requires breakdown of a stated hypothesis.
\item The selection persists under the uniform effective bound of Theorem~\ref{thm:stability} and is compatible with locally covariant free CAR quantization, subject to the signed antiparticle charge of Theorem~\ref{thm:fock}.
\end{enumerate}
\end{theorem}
\begin{proof}
Claims (1) and (2) are Theorems~\ref{thm:reduction}, \ref{thm:orientation} and \ref{thm:spin}. Claim (3) follows from Lemmas~\ref{lem:current} and \ref{lem:unique} and depends on the tensorial and restricted-operator clauses of EDG2$'$. Claim (4) uses Eq.~\eqref{eq:cliff-identity}. Claim (5) is Theorem~\ref{thm:selection}, or Proposition~\ref{prop:convex} with its different closure hypothesis. Claim (6) is Theorem~\ref{thm:parity} and Proposition~\ref{prop:wall}. Claim (7) is Theorem~\ref{thm:stability} together with the standard CAR construction and Theorem~\ref{thm:fock}.
\end{proof}

\begin{table*}[t]
\caption{Logical status of the inputs and the conclusions. A theorem status refers to the precise assumptions in the same row and does not remove the independent premises.}\label{tab:status}
\centering\small
\begin{tabular}{@{}>{\raggedright\arraybackslash}p{0.265\textwidth}>{\raggedright\arraybackslash}p{0.41\textwidth}>{\raggedright\arraybackslash}p{0.245\textwidth}@{}}
\toprule
Statement & Required input & Logical status \\
\midrule
Four-dimensional Lorentz sector & Empirical geometry $(M,g,u)$ & Physical input \\
Hamiltonian accessibility & Unitary smooth accessible procedures and HCB & Conditional operational bridge \\
Reduced symplecticity & Closed $\Omega$, HCB, regular smooth quotient & Theorem \\
Global orientation & Ray-space QOC, quotient gluing, four reduced dimensions & Theorem under observer-atlas hypotheses \\
Global spin structure & GH, time orientation, derived spatial orientation & Theorem in this branch \\
Smooth surface three-form & Tensorial EDG2$'$ density & Representation theorem; tensoriality assumed \\
Clifford three-form uniqueness & Quadratic leading term and specified pre-Hodge grade-three class & Restricted natural-map theorem \\
Spin-response interpretation & Variational Dirac spin connection & Standard-field-theory motivation \\
Single Weyl carrier & Causal flux sign, nonzero $\kappa$, ELC or convex alternative & Conditional selection theorem \\
Absolute external left/right label & Same parity-covariant premises & Underdetermined \\
Time-monotone accumulated record & Extra source/balance law, boundary control & Not derived \\
Weak gauge charges and anomaly cancellation & Independent interacting-field input & Outside the theorem \\
\bottomrule
\end{tabular}
\end{table*}

\section{Discussion and limitations}\label{sec:discussion}
\subsection{What is explained, and what remains open}
The result identifies a set of geometric and causal conditions sufficient to reduce a Dirac representation to one Lorentz-Weyl module. A symplectic-square orientation fixes the global chirality operator; a locally tensorial spin-response record gives an orientation-odd axial flux; and closure of elementary amplitudes turns a flux-sign constraint into a carrier-selection theorem. The sign result is explicit for both signs of the nonzero record coefficient and the counterexamples delimit the necessity of its operator class and carrier closure.

The main unresolved physical step is the operational origin of the surface-record law. A differentiable irreversible counting process need not define a tensorially linear surface current. Even when such a current exists, standard Dirac theory admits several local bilinears, and the usual positive vector current does not select a chirality. The identification of a distinguishability record with the antisymmetric Dirac spin response is independently motivated by the variational coupling in Eq.~\eqref{eq:spin-current}; its experimental status and possible dynamical derivation remain open. The distinct condition of positivity refers to elementary throughput. It does not prove positive entropy production, as Eq.~\eqref{eq:balance} makes explicit.

The coherent linear-carrier assumption ELC also remains independent. Quantum amplitude superposition motivates it for a fixed elementary field representation, while the mixed-spinor example demonstrates the failure of selection from positivity alone. A convex-amplitude closure provides a second sufficient condition with its own physical scope. Neither assumption follows from arbitrary probabilistic mixtures.

The relative-orientation construction cannot decide which externally labelled Weyl representation Nature realizes. The model pair in Section~\ref{sec:parity} establishes this underdetermination. Adding a fixed parity-odd calibration would select a label, but it would be extra physical input. The result is therefore separate from spontaneous parity breaking and from the dynamical emergence of the electroweak representation spectrum.

\subsection{Relation to chiral dynamics and model building}
The construction acts before an internal gauge representation is specified. Conventional Lorentz and quantum-field-theory machinery governs chiral oscillations \cite{epjc_blasone2025}, spinor bilinear classifications \cite{epjc_lee2024}, spin topology \cite{epjc_dossantos2026}, and deconstructed chiral localization \cite{epjc_singh2026}. Left--right gauge theories generate parity violation through a different symmetry-breaking problem \cite{ref26}. Spectral noncommutative geometry investigates gauge and Higgs data in a more extensive algebraic framework \cite{ref27,ref28}. Our theorem does not compete with those dynamical completions. It isolates a kinematic ingredient whose subsequent coupling to gauge fields must be supplied or derived separately.

A phenomenological extension would require a gauge connection acting on the selected module, a consistent anomaly-free set of additional fields, and dynamical equations for the record and gauge interactions. The present construction admits antiparticles through standard CAR quantization. It neither removes opposite-chirality fields in other internal representations nor supplies a charge-conjugation or CP-violating observable. In its current form it establishes a conditional structural statement, without a parameter-free observable prediction of weak-interaction parity violation.

\subsection{Programme status and concrete next steps}\label{subsec:programme}
This study is intended as an initial structural contribution in a wider investigation of operational distinguishability and chiral matter. Describing it as the first stage is not evidence for an unproved premise and does not relax the present logical burden. The present result is complete only as a conditional theorem; the following tasks are separate research questions rather than conclusions established here.

First, construct a specified microscopic instrument or open-system model whose locally calibrated spin-sensitive output admits a smooth tensorial three-form and reproduces the Clifford-grade-three response without post-selecting the chiral subspace. Second, derive or independently verify closure of its elementary amplitude preparations, rather than invoking convex mixtures of density matrices. Third, check the required sign for all locally realizable configurations and distinguish transport positivity from an actual irreversible source term. Fourth, identify a concrete anomaly-consistent matter and gauge sector and test whether the geometric selection criterion constrains its couplings or measurable parity asymmetry. The current paper supplies no numerical values or predictions from these later steps; they cannot be used to establish novelty or empirical validation of the present theorem.

\subsection{Prospects for a more primitive derivation}
The next step in this research programme would be an independently calibrated realization. A route toward a foundational derivation would proceed by independently characterizing the continuum limit of accessible quantum instruments, constructing a local signed record as a geometric current, identifying its spin-coupling channel, and establishing a coherent closure rule for elementary amplitudes. Each step has a distinct mathematical obstruction. HCB excludes general dissipative generators; RCR requires a regular quotient and a verifiable rank-to-geometry bridge \cite{tulupov2026}; QOC requires full transition-probability covariance; EDG2$'$ requires tensorial surface locality and a restricted leading operator; and ELC requires complex-linear amplitude closure. The present manuscript identifies these requirements as explicit research targets. Their joint realization would connect the selection theorem to a physical measurement construction.

\section*{Statements and declarations}
\noindent\textbf{Funding.} This research did not receive any specific grant from public, commercial or not-for-profit funding agencies.

\noindent\textbf{Competing interests.} The authors declare no competing financial or non-financial interests relevant to the content of this article.

\noindent\textbf{Author contributions.} A.S.O. developed the detailed derivations and prepared the manuscript. A.M.T. contributed to conceptual framing and critical revision. Both authors discussed the results and approved the manuscript.

\noindent\textbf{Data availability.} No datasets were generated or analysed. The work is theoretical.

\noindent\textbf{Code availability.} No software or computational dataset is needed to reproduce the stated analytic results.

\noindent\textbf{Ethics approval and consent to participate.} Not applicable.

\noindent\textbf{Consent for publication.} Not applicable.

\noindent\textbf{Use of AI-assisted tools.} During manuscript preparation, generative AI tools, including ChatGPT (OpenAI), were used for substantive language revision, LaTeX restructuring, exploratory checks of mathematical arguments and bibliography consistency. Such assistance is not a substitute for independent derivation or peer review. The authors retain responsibility for the mathematical arguments, physical interpretations, references, and final content. 

\appendix
\section{Clifford and flux conventions}\label{app:clifford}
We use signature $(+---)$, anticommutation $\{\gamma^\mu,\gamma^\nu\}=2g^{\mu\nu}$ and $\bar\psi=\psi^\dagger\gamma^0$. The symbol $\gamma_{\mu_1\ldots\mu_r}$ means the fully antisymmetrized product with weight $1/r!$. Fix an oriented orthonormal frame in which $\eps_{0123}=+1$ and $\gamma^5=i\gamma^0\gamma^1\gamma^2\gamma^3$. Then
\begin{equation}\label{eq:appendix-a}
 i\gamma_{\mu\nu\rho}=\eps_{\mu\nu\rho\sigma}\gamma^\sigma\gamma^5.
\end{equation}
For the test component $\mu\nu\rho=123$,
\begin{equation}\label{eq:appendix-test}
 i\gamma_{123}=-i\gamma^1\gamma^2\gamma^3=-\gamma^0\gamma^5=\eps_{1230}\gamma^0\gamma^5.
\end{equation}
This component verifies the sign without appealing to an adjustable normalization. With the Hodge-dual definition $j=\iota_\Xi\eps$, it implies the minus sign in Eq.~\eqref{eq:axial}.

The Dirac adjoint of a triple Clifford product is anti-self-adjoint, so $i\bar\psi\gamma_{\mu\nu\rho}\psi$ is real. The projectors obey
\begin{equation}\label{eq:appendix-projectors}
 \bar\psi_L=\bar\psi P_R,\qquad\bar\psi_R=\bar\psi P_L.
\end{equation}
These identities imply the even-Clifford vanishing statement in Lemma~\ref{lem:even}. The positive Hermitian pairing in Eq.~\eqref{eq:positive-pairing} is represented by $\varphi^\dagger\psi$ in the rest frame of the future unit normal.

\section{The surface-current hypothesis and a normal-dependent counterexample}\label{app:currents}
Order-zero absolute continuity for a functional on embedded hypersurfaces says that its local density can be integrated against the induced positive hypersurface measure. Without tensorial linearity, it can have the generic form
\begin{equation}\label{eq:surface-anisotropy}
 F[\Sigma]=\int_\Sigma f(x,n_\Sigma,\psi)\,\D\Sigma,
\end{equation}
where $f$ depends nonlinearly on the future normal $n_\Sigma$. The functional remains additive over disjoint patches. A reference vector $a$ gives a simple illustrative choice $f=(a\cdot n_\Sigma)^3$, together with the prescribed orientation sign. It is generally not the evaluation of a three-form because its normal dependence is nonlinear. The example uses an external vector and is meant to show why additivity alone is insufficient; it is not an admissible leading tensor-minimal EDG2$'$ record.

The tensorial clause of EDG2$'$ separately requires an alternating trilinear density $T_\psi(v_1,v_2,v_3)$, which immediately corresponds to a smooth three-form. Under this condition the Radon--Nikodym density on a spacelike patch agrees with the contraction $n_\mu\Xi^\mu$ after the orientation is fixed. For general current theory and differential-form conventions see Ref.~\cite{ref15}.

\section{Quantum-realized parity-related observer models}\label{app:models}
The constant-form witnesses of Section~\ref{sec:parity} already prove model-theoretic parity underdetermination. Here we exhibit a distinct quantum realization with nonidentity observer transitions; it is not identified with the constant-form model by a claimed global symplectomorphism.

Topologically identify $M=\mathbb R^{1,3}$ with the affine chart $\mathbb C^2$ of $\mathbb{CP}^1\times\mathbb{CP}^1$, using coordinates $z_1,z_2$. Give the same underlying manifold the flat Lorentz metric and future orientation of Section~\ref{sec:parity}. On each projective factor the Fubini--Study two-form is
\begin{equation}\label{eq:fsfactor}
 \omega_j=\frac{i\,\D z_j\wedge\D\bar z_j}{(1+|z_j|^2)^2},\qquad j=1,2,
\end{equation}
up to a common positive normalization. Consider the two four-dimensional quantum pullback forms
\begin{equation}\label{eq:fsopposite}
 \Omega_A^q=\omega_1+\omega_2,\qquad
 \Omega_B^q=\omega_1-\omega_2.
\end{equation}
Their half-squares are $\omega_1\wedge\omega_2$ and $-\omega_1\wedge\omega_2$. Each form is closed and nondegenerate on $\mathbb C^2$.

Let $\sigma_A$ be the Segre embedding of product rays into $\mathbb{CP}^3$ and let $\sigma_B$ compose it with complex conjugation on the second factor:
\begin{equation}\label{eq:segre}
 \sigma_A([\xi],[\eta])=[\xi\otimes\eta],\qquad
 \sigma_B([\xi],[\eta])=[\xi\otimes\overline\eta].
\end{equation}
Pullback of the ambient Fubini--Study form gives the two forms in Eq.~\eqref{eq:fsopposite}, after choosing common normalization. This displays a direct ray-space origin of their opposite symplectic orientations.

Choose nonidentity phase rotations $U_1,U_2$ on the two qubit factors, with charts chosen so that they overlap. In Model A the induced ambient unitary transition is $U_1\otimes U_2$. In Model B, with the conjugated second factor, it is $U_1\otimes\overline{U_2}$. They preserve transition probabilities and the corresponding pullback forms. Thus QOC acts nontrivially in each model, while the two reduced symplectic squares have opposite orientations relative to one external reference. Take the same Clifford three-form law in both models and choose the appropriate selected carrier according to Theorem~\ref{thm:selection}. This realizes the same underdetermination in a nontrivial quantum observer atlas.

\section{Fock-space charge in the stationary regime}\label{app:fock}
For a selected free massless Weyl field, the normal-ordered stationary signed charge is
\begin{equation}\label{eq:appendix-fock}
 \hat Q_{\DG}=|\kappa|\sum_r\int\D\Pi_p\,[a_r^\dagger(p)a_r(p)-b_r^\dagger(p)b_r(p)].
\end{equation}
On the occupation-number basis it acts as
\begin{equation}\label{eq:occupation}
 \hat Q_{\DG}|n,m\rangle=|\kappa|(n-m)|n,m\rangle.
\end{equation}
The spectrum is unbounded above and below. This agrees with the ordinary signed charge of a complex fermion and shows why the elementary classical record positivity condition cannot be interpreted as positive total quantum charge on Fock space.

\setlength{\bibsep}{0.8em}
\bibliography{references}

@misc{ref1,
  author       = {Tishin, Alexander M.},
  title        = {Distinguishability Geometry as a Universal Foundation for Physical Dynamics},
  year         = {2026},
  howpublished = {Research Square preprint},
  note      = {version 4},
  doi       = {10.21203/rs.3.rs-10001438/v4}
}

@misc{ref2,
  author       = {Tishin, Alexander M.},
  title        = {Before Geometry: Four-Dimensionality and Quantum Entanglement from Operational Distinguishability},
  year         = {2026},
  howpublished = {Zenodo preprint},
  doi     = {10.5281/zenodo.21159016}
}

@article{ref3,
  author  = {Weyl, Hermann},
  title   = {Elektron und Gravitation},
  journal = {Z. Phys.},
  volume  = {56},
  pages   = {330--352},
  year    = {1929},
  doi     = {10.1007/BF01339504}
}

@article{ref4,
  author  = {Wigner, Eugene P.},
  title   = {On Unitary Representations of the Inhomogeneous Lorentz Group},
  journal = {Ann. Math.},
  volume  = {40},
  pages   = {149--204},
  year    = {1939},
  doi     = {10.2307/1968551}
}

@book{ref5,
  author    = {Weinberg, Steven},
  title     = {The Quantum Theory of Fields, Vol. I: Foundations},
  publisher = {Cambridge University Press},
  address   = {Cambridge},
  year      = {1995},
  doi     = {10.1017/CBO9781139644167}
}

@article{ref6,
  author  = {Pal, Palash B.},
  title   = {Dirac, Majorana and Weyl Fermions},
  journal = {Am. J. Phys.},
  volume  = {79},
  pages   = {485--498},
  year    = {2011},
  doi     = {10.1119/1.3549729}
}

@article{ref7,
  author  = {Dreiner, Herbi K. and Haber, Howard E. and Martin, Stephen P.},
  title   = {Two-component Spinor Techniques and Feynman Rules for Quantum Field Theory and Supersymmetry},
  journal = {Phys. Rep.},
  volume  = {494},
  pages   = {1--196},
  year    = {2010},
  doi     = {10.1016/j.physrep.2010.05.002}
}

@article{ref8,
  author  = {Lee, T. D. and Yang, C. N.},
  title   = {Question of Parity Conservation in Weak Interactions},
  journal = {Phys. Rev.},
  volume  = {104},
  pages   = {254--258},
  year    = {1956},
  doi     = {10.1103/PhysRev.104.254}
}

@article{ref9,
  author  = {Wu, C. S. and Ambler, E. and Hayward, R. W. and Hoppes, D. D. and Hudson, R. P.},
  title   = {Experimental Test of Parity Conservation in Beta Decay},
  journal = {Phys. Rev.},
  volume  = {105},
  pages   = {1413--1415},
  year    = {1957},
  doi     = {10.1103/PhysRev.105.1413}
}

@book{ref10,
  author    = {{Cannas da Silva}, Ana},
  title     = {Lectures on Symplectic Geometry},
  publisher = {Springer},
  address   = {Berlin},
  year      = {2001},
  doi     = {10.1007/978-3-540-45330-7}
}

@book{ref11,
  author    = {Lawson, H. Blaine and Michelsohn, Marie-Louise},
  title     = {Spin Geometry},
  publisher = {Princeton University Press},
  address   = {Princeton, NJ},
  year      = {1989}
}

@book{ref12,
  author    = {Penrose, Roger and Rindler, Wolfgang},
  title     = {Spinors and Space-Time, Vol. 1: Two-Spinor Calculus and Relativistic Fields},
  publisher = {Cambridge University Press},
  address   = {Cambridge},
  year      = {1984},
  doi     = {10.1017/CBO9780511564048}
}

@article{ref13,
  author  = {Bernal, Antonio N. and S{\'a}nchez, Miguel},
  title   = {On Smooth Cauchy Hypersurfaces and Geroch's Splitting Theorem},
  journal = {Commun. Math. Phys.},
  volume  = {243},
  pages   = {461--470},
  year    = {2003},
  doi     = {10.1007/s00220-003-0982-6}
}

@article{ref14,
  author  = {Geroch, Robert P.},
  title   = {Spinor Structure of Space-Times in General Relativity. I},
  journal = {J. Math. Phys.},
  volume  = {9},
  pages   = {1739--1744},
  year    = {1968},
  doi     = {10.1063/1.1664507}
}

@book{ref15,
  author    = {{de Rham}, Georges},
  title     = {Differentiable Manifolds: Forms, Currents, Harmonic Forms},
  publisher = {Springer},
  address   = {Berlin},
  year      = {1984}
}

@article{ref16,
  author  = {Dimock, Jonathan},
  title   = {Dirac Quantum Fields on a Manifold},
  journal = {Trans. Am. Math. Soc.},
  volume  = {269},
  pages   = {133--147},
  year    = {1982},
  doi     = {10.1090/S0002-9947-1982-0637032-8}
}

@article{ref17,
  author  = {Brunetti, Romeo and Fredenhagen, Klaus and Verch, Rainer},
  title   = {The Generally Covariant Locality Principle},
  journal = {Commun. Math. Phys.},
  volume  = {237},
  pages   = {31--68},
  year    = {2003},
  doi     = {10.1007/s00220-003-0815-7}
}

@article{ref18,
  author  = {Sanders, Ko},
  title   = {The Locally Covariant Dirac Field},
  journal = {Rev. Math. Phys.},
  volume  = {22},
  pages   = {381--430},
  year    = {2010},
  doi     = {10.1142/S0129055X10003990}
}

@article{ref19,
  author  = {Dappiaggi, Claudio and Hack, Thomas-Paul and Pinamonti, Nicola},
  title   = {The Extended Algebra of Observables for Dirac Fields and the Trace Anomaly of Their Stress-Energy Tensor},
  journal = {Rev. Math. Phys.},
  volume  = {21},
  pages   = {1241--1312},
  year    = {2009},
  doi     = {10.1142/S0129055X09003717}
}

@article{ref20,
  author  = {Zahn, Jochen},
  title   = {The Renormalized Locally Covariant Dirac Field},
  journal = {Rev. Math. Phys.},
  volume  = {26},
  pages   = {1330012},
  year    = {2014},
  doi     = {10.1142/S0129055X13300124}
}

@article{ref24,
  author  = {Takahashi, F. and others},
  collaboration = {Particle Data Group},
  title   = {Review of Particle Physics},
  journal = {Int. J. Mod. Phys. A},
  volume  = {41},
  pages   = {2630011},
  year    = {2026},
  doi     = {10.1142/S0217751X26300115}
}

@article{ref25,
  author  = {Feynman, Richard P. and Gell-Mann, Murray},
  title   = {Theory of the Fermi Interaction},
  journal = {Phys. Rev.},
  volume  = {109},
  pages   = {193--198},
  year    = {1958},
  doi     = {10.1103/PhysRev.109.193}
}

@article{ref26,
  author  = {Senjanovi{\'c}, Goran and Mohapatra, Rabindra N.},
  title   = {Exact Left-Right Symmetry and Spontaneous Violation of Parity},
  journal = {Phys. Rev. D},
  volume  = {12},
  pages   = {1502--1505},
  year    = {1975},
  doi     = {10.1103/PhysRevD.12.1502}
}

@article{ref27,
  author  = {Chamseddine, Ali H. and Connes, Alain and Marcolli, Matilde},
  title   = {Gravity and the Standard Model with Neutrino Mixing},
  journal = {Adv. Theor. Math. Phys.},
  volume  = {11},
  pages   = {991--1089},
  year    = {2007},
  doi     = {10.4310/ATMP.2007.v11.n6.a3}
}

@article{ref28,
  author  = {Devastato, Agostino and Kurkov, Maxim and Lizzi, Fedele},
  title   = {Spectral Noncommutative Geometry, Standard Model and All That},
  journal = {Int. J. Mod. Phys. A},
  volume  = {34},
  pages   = {1930010},
  year    = {2019},
  doi     = {10.1142/S0217751X19300102}
}

@article{mir2026,
  author  = {Mir, Sameer Ahmad and Shabir, Arshid and Tiwari, Swatantra Kumar and Faizal, Mir},
  title   = {Geometric Baryogenesis With Chiral-Time Equivalence},
  journal = {Fortschritte der Physik},
  volume  = {74},
  number  = {3},
  pages   = {e70088},
  year    = {2026},
  doi     = {10.1002/prop.70088}
}

@article{finster2024,
  author  = {Finster, Felix and Gresnigt, Niels G. and Isidro, Jos{\'e} M. and Marcian{\`o}, Antonino and Paganini, Claudio F. and Singh, Tejinder P.},
  title   = {Causal Fermion Systems and Octonions},
  journal = {Fortschritte der Physik},
  volume  = {72},
  number  = {11},
  pages   = {2400055},
  year    = {2024},
  doi     = {10.1002/prop.202400055}
}

@article{darianoperinotti2017,
  author  = {D'Ariano, Giacomo Mauro and Perinotti, Paolo},
  title   = {Quantum Cellular Automata and Free Quantum Field Theory},
  journal = {Frontiers of Physics},
  volume  = {12},
  number  = {1},
  pages   = {120301},
  year    = {2017},
  doi     = {10.1007/s11467-016-0616-z}
}

@article{baptista2026,
  author  = {Baptista, Jo{\~a}o},
  title   = {Chiral Interactions of Fermions and Massive Gauge Fields in Kaluza--Klein Models},
  journal = {JHEP},
  volume  = {05},
  pages   = {008},
  year    = {2026},
  doi     = {10.1007/JHEP05(2026)008}
}

@article{larue2023,
  author  = {Larue, R{\'e}my and Quevillon, J{\'e}r{\'e}mie and Zwicky, Roman},
  title   = {Trace Anomaly of Weyl Fermions via the Path Integral},
  journal = {JHEP},
  volume  = {12},
  pages   = {064},
  year    = {2023},
  doi     = {10.1007/JHEP12(2023)064}
}

@article{kamada2025,
  author  = {Kamada, Kohei and Kume, Jun'ya},
  title   = {On the Inefficiency of Fermion Level-Crossing under the Parity-Violating Spin-2 Gravitational Field},
  journal = {JHEP},
  volume  = {06},
  pages   = {203},
  year    = {2025},
  doi     = {10.1007/JHEP06(2025)203}
}

@article{maleknejad2025,
  author  = {Maleknejad, Azadeh and Kopp, Joachim},
  title   = {Weyl Fermion Creation by Cosmological Gravitational Wave Background at 1-Loop},
  journal = {JHEP},
  volume  = {01},
  pages   = {023},
  year    = {2025},
  doi     = {10.1007/JHEP01(2025)023}
}

@article{epjc_lee2024,
  author  = {Lee, Cheng-Yang},
  title   = {Equivalence of regular spinor fields},
  journal = {Eur. Phys. J. C},
  volume  = {84},
  pages   = {943},
  year    = {2024},
  doi     = {10.1140/epjc/s10052-024-13319-y}
}

@article{epjc_blasone2025,
  author  = {Blasone, Massimo and Giacosa, Francesco and Smaldone, Luca and Torrieri, Giorgio},
  title   = {Chiral oscillations in finite time quantum field theory},
  journal = {Eur. Phys. J. C},
  volume  = {85},
  pages   = {523},
  year    = {2025},
  doi     = {10.1140/epjc/s10052-025-14165-2}
}

@article{epjc_dossantos2026,
  author  = {{dos Santos Filho}, Lu{\'i}s R.},
  title   = {Effective field description of vector and tensorial couplings in exotic spinor dynamics: a geometrical approach to Rashba-like interactions},
  journal = {Eur. Phys. J. C},
  volume  = {86},
  pages   = {495},
  year    = {2026},
  doi     = {10.1140/epjc/s10052-026-15723-y}
}

@article{epjc_singh2026,
  author  = {Singh, Aadarsh},
  title   = {Independent chiral control in theory-space models: a rank-preserving framework and its application to neutrino mass generation},
  journal = {Eur. Phys. J. C},
  volume  = {86},
  pages   = {743},
  year    = {2026},
  doi     = {10.1140/epjc/s10052-026-15918-3}
}

@misc{tulupov2026,
  author = {Tulupov, Nikolay D. and Tishin, Alexander M.},
  title = {Compositional Operational Rank, Smooth Realizability, and Dimensional Reduction},
  year = {2026},
  howpublished = {arXiv preprint arXiv:2610.35764},
  eprint = {2610.35764},
  archivePrefix = {arXiv},
  doi = {10.48550/arXiv.2610.35764}
}

@misc{ashtekarSchilling1997,
  author = {Ashtekar, Abhay and Schilling, Troy A.},
  title = {Geometrical Formulation of Quantum Mechanics},
  year = {1997},
  howpublished = {arXiv preprint gr-qc/9706069},
  url = {https://arxiv.org/abs/gr-qc/9706069},
  eprint = {gr-qc/9706069}
}

@article{bargmann1964,
  author = {Bargmann, Valentine},
  title = {Note on {Wigner}'s Theorem on Symmetry Operations},
  journal = {Journal of Mathematical Physics},
  volume = {5},
  pages = {862--868},
  year = {1964},
  doi = {10.1063/1.1704188}
}

@article{hehl1976,
  author = {Hehl, Friedrich W. and von der Heyde, Paul and Kerlick, G. David and Nester, James M.},
  title = {General Relativity with Spin and Torsion: Foundations and Prospects},
  journal = {Reviews of Modern Physics},
  volume = {48},
  pages = {393--416},
  year = {1976},
  doi = {10.1103/RevModPhys.48.393}
}

@article{shapiro2002,
  author = {Shapiro, Ilya L.},
  title = {Physical Aspects of the Space-Time Torsion},
  journal = {Physics Reports},
  volume = {357},
  pages = {113--213},
  year = {2002},
  doi = {10.1016/S0370-1573(01)00030-8}
}
\end{document}